\documentclass{emulateapj}
\usepackage{apjfonts}
\usepackage{epsfig}

\usepackage[latin1]{inputenc}
\usepackage{lscape}

\shorttitle{Triple System in XZ~Tau}
\shortauthors{Carrasco-Gonz\'alez et al.}

\slugcomment{The Astrophysical Journal Letters}

\received{November 14, 2008}
\accepted{January 22, 2009}

\begin{document}

\title{High Angular Resolution Radio Observations of the HL/XZ~Tau Region: Mapping the 50 AU Protoplanetary Disk around HL~Tau and Resolving XZ~Tau~S into a 13 AU Binary}

\author{Carlos~Carrasco-Gonz\'alez\altaffilmark{1,2}, Luis~F.~Rodr\'{\i}guez\altaffilmark{2}, Guillem~Anglada\altaffilmark{1}, Salvador~Curiel\altaffilmark{3}}

\altaffiltext{1}{Instituto de Astrof\'{\i}sica de Andaluc\'{\i}a (CSIC), Camino
Bajo de Hu\'etor 50, E-18008 Granada, Spain; charly@iaa.es,
guillem@iaa.es}

\altaffiltext{2}{Centro de Radioastronom\'{\i}a y Astrof\'{\i}sica (UNAM),
Apartado Postal 3-72 (Xangari), 58089 Morelia, Michoac\'an, M\'exico;
l.rodriguez@astrosmo.unam.mx}

\altaffiltext{3}{Instituto de Astronom\'{i}a (UNAM), Apartado Postal 70-264, D.F. 04510, Mexico; scuriel@astroscu.unam.mx }

\begin{abstract}

We present new 7 mm and archive 1.3 cm high angular resolution observations of the HL/XZ~Tau region made with the VLA. At 7 mm, the emission from HL~Tau seems to be arising in a clumpy disk with radius of order 25 AU. The 1.3 cm emission from XZ~Tau shows the emission from a binary system with 0$\farcs$3 (42 AU) separation, known from previous optical/IR observations. However, at 7 mm, the southern radio component resolves into a binary with 0$\farcs$09 (13 AU) separation, suggesting that XZ~Tau is actually a triple star system. We suggest that the remarkable ejection of gas from the XZ~Tau system observed with the HST may be related to a periastron passage of this newly discovered close binary system.

\end{abstract}

\keywords{ISM: individual (XZ~Tau, HL~Tau) --- ISM: jets and outflows --- radio 
continuum: ISM --- stars: formation}

\section{Introduction}

 The region encompassing the stars HL/XZ~Tau, in the northeastern part of the L1551 dark cloud, has been the subject of many studies over the years. This region lies at a distance of 140 pc (e.g., Kenyon et al. 1994; Torres et al. 2009) and it is particularly rich in HH jets, being one of the regions where this phenomenon was first identified (Mundt \& Fried 1983).

 HL~Tau is one of the most intensively studied T Tauri stars. Since the first proposal that this star is associated with a nearly edge-on circumstellar disk (Cohen 1983), numerous studies have been carried out in order to image the proposed disk (e.g., Sargent \& Beckwith 1991; Wilner et al. 1996; Looney et al. 2000). This star has been proposed as the exciting source of a molecular outflow (e.g., Torrelles et al. 1987; Monin et al. 1996). HL~Tau is also the source of a collimated optical jet-counterjet system that has been extensively studied (e.g., Mundt et al. 1990; Rodr\'{i}guez et al. 1994; Anglada et al. 2007). Very recently, Greaves et al. (2008) presented high angular resolution maps at 1.3 cm of the jet-disk system in HL~Tau. These authors reported the detection of a weak 1.3 cm source which they propose is a 14 M$_J$ proto-planet orbiting at a radius of $\sim$65 AU around HL~Tau. 

 XZ~Tau, located $\sim$25$\arcsec$ to the east of HL~Tau, is a close binary system composed of a T Tauri star and a cool companion separated by 0$\farcs$3 (Haas et al. 1990). XZ~Tau is also the source of an optical outflow, as revealed, e.g., by the studies of Mundt et al. (1990). The spectacular sequence of Hubble Space Telescope (HST) images of Krist et al. (1999) shows evidence of the expansion of nebular emission, moving away from XZ~Tau with a velocity of $\sim$70 km s$^{-1}$. This bubble of nebular emission is different from the collimated jets seen in other young stars, for example HL~Tau. Recent new HST observations (Krist et al. 2008) revealed a succession of bubbles and a fainter counterbubble. In addition, they reveal that both components of the XZ~ Tau binary are also driving collimated jets. Krist et al. (2008) proposed that the bubble is the result of a large velocity pulse in the collimated jet driven by the southern component of the XZ~Tau system.
 
 In this Letter, we present the results of sensitive high-angular resolution Very Large Array (VLA) observations at 1.3 cm and 7 mm of HL/XZ~Tau. These observations allowed us to map the dust emission from the disk of HL~Tau at scales of $\sim$7 AU and reveal that XZ~Tau is actually a triple-star system with the southern optical component resolved into a close binary with a separation of only 13 AU.

\section{Observations}

 Observations at 7 mm continuum were made using the VLA of the National Radio Astronomy Observatory (NRAO)\footnote{The NRAO is a facility of the National Science Foundation operated under cooperative agreement by  Associated Universities, Inc.}\ . The data were taken in A, B and D configurations (VLA Project Codes: AC0763, AC0850 and AC0816). Total on-source integration times were $\sim$3~h at A Configuration (observed in three runs during 2004 October 15, 17 and 26), $\sim$1~h at B Configuration (2006 September 19), and $\sim$1.5~h at D Configuration (observed in two runs during 2006 January 20 and 22). The fast-switching technique was used for all the observations. Phase and flux calibrators were 0431+175 and 3C286, respectively. The same phase center was used for all the observations. 

 We also used archive continuum data at 1.3~cm taken with the VLA in its A Configuration (VLA Project Code: AG0711) with a total on-source integration time of $\sim$12~h (observed in two runs during 2006 March 12 and April 1). Phase and flux calibrators were 0431+206 and 3C286, respectively. The Pie Town antenna was used originally in these observations, but we did not include it in our analysis. 

\begin{figure*}  
\begin{center}

\includegraphics[width=\textwidth]{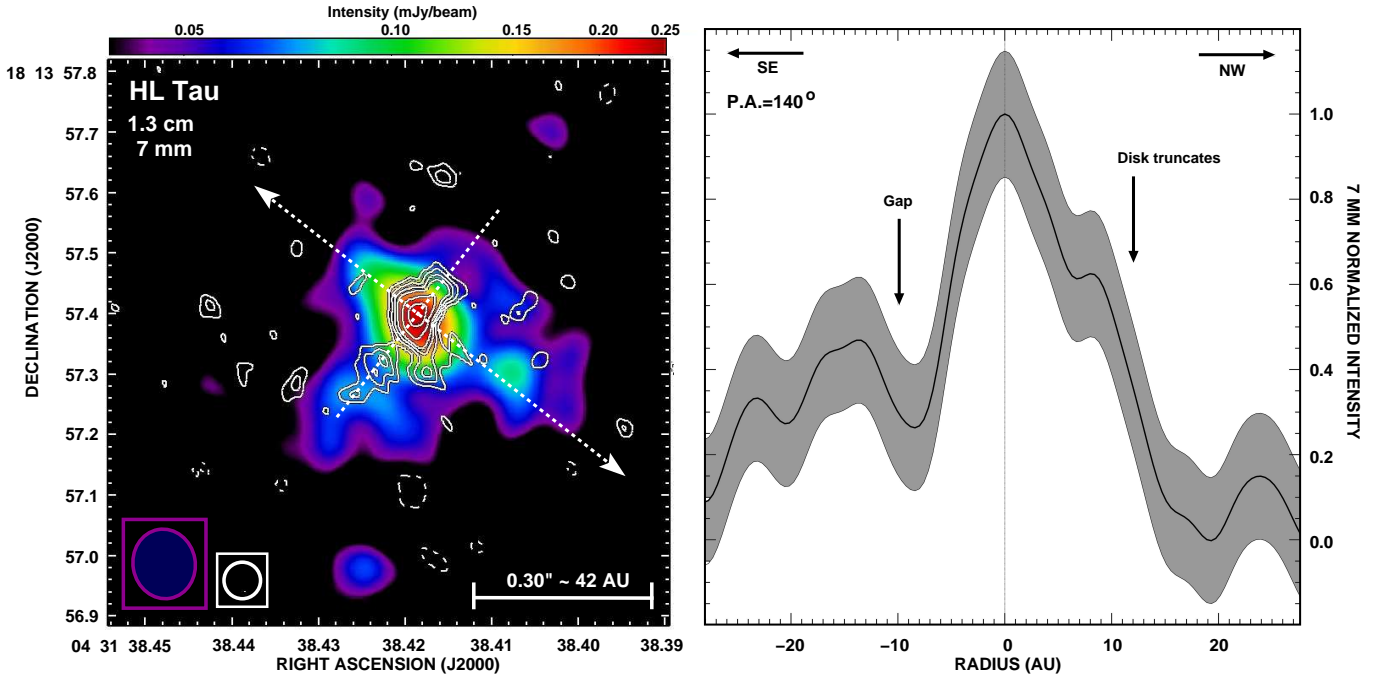}

\caption{\footnotesize{\textbf{(a)} Superposition of the 7 mm map (contours) over the 1.3 cm continuum map (color scale) of HL~Tau. Contours at 7 mm are $-$3, 3, 4, 5, 6, 8, 10, 12, 14, 18, 22, 26 and 32 times the rms of the map, 50 $\mu$Jy. The dashed lines mark the directions of the optical jet (P.A.=50$^\circ$) and the disk (P.A.=140$^\circ$). The half power contour of the 1.3 cm (filled ellipse) and 7 mm (empty ellipse) beams are shown in the bottom left corner. \textbf{(b)} A 7 mm intensity profile along the major axis of the disk (P.A.=140$^\circ$), obtained by averaging the emission at $\pm$0$\farcs$05 from the axis marked in (a). The intensity error is indicated by the thickness of the profile. The disk truncates at a radius of $\sim$12 AU to the NW, and shows a gap at a radius of $\sim$10 AU to the SE. We speculate that these features could be related to the formation of a proto-planet (see text).}}

\label{FigHL}  
\end{center}
\end{figure*}

 Data editing and calibration were carried out using the Astronomical Image Processing System (AIPS) package of NRAO, following the standard high-frequency VLA procedures. Primary beam corrections were applied using the task PBCOR of AIPS. 
  
 Maps at 7 mm were made by concatenating the data from all the configurations (A, B and D). This allowed us to improve by $\sim$20\% the signal-to-noise ratio of the A configuration data without significant detriment of the angular resolution. The 7 mm map of HL~Tau shown here was made using data with baselines $>$150 k$\lambda$ in order to resolve out extended emission that surrounds HL~Tau. The nature of this extended emission is not fully understood, but probably originates from diffuse dust in the region. The 1.3 cm map of HL~Tau shown in Greaves et al. (2008) was obtained from the same data, and, as expected, is very similar to our 1.3 cm map. 
 
 The rms noises of the natural weighting maps were 13~$\mu$Jy at 1.3 cm, and 50~$\mu$Jy at 7 mm. Synthesized beams are $\sim$0$\farcs$10 at 1.3 cm and $\sim$0$\farcs$05 at 7 mm. Bandwidth smearing at the position of XZ~Tau degrades the resolution by $\sim$5\% in the radial direction, which does not affect our analysis.
 
 This region has a known proper motion of 0$\farcs$025$\pm$0$\farcs$0035 yr$^{-1}$ (Rodr\'{i}guez et al. 2003), so we shifted the 2006 1.3 cm data by 0$\farcs$04 to align them with the 2004 7 mm data (with uncertainties of 0$\farcs$005 in proper motions and 0$\farcs$01 in absolute astrometry at each epoch). 

\begin{figure*}  
\begin{center}

\includegraphics[width=1.\textwidth]{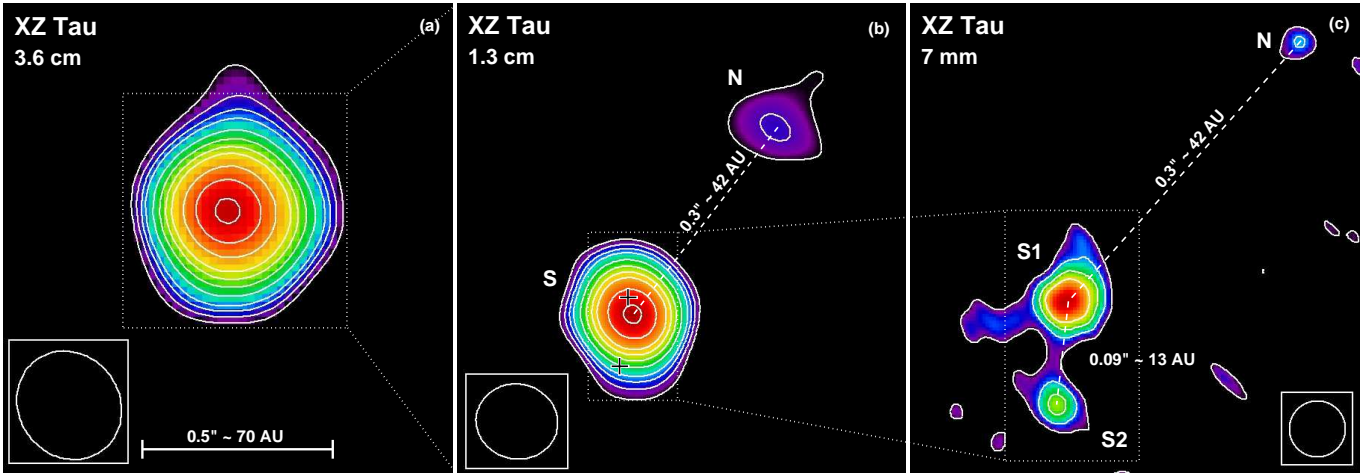}

\caption{\footnotesize{Radio continuum maps of XZ~Tau at 3.6 cm (a), 1.3 cm (b) and 7 mm (c). The contours are $-$3, 3, 4, 5, 6, 8, 10, 12, 14, 18, 22 and 26 times the rms of each map, 7 $\mu$Jy (3.6 cm), 13 $\mu$Jy (1.3 cm) and 50 $\mu$Jy (7 mm). At 3.6 cm a single radio source is detected. The 1.3 cm map shows two continuum sources, labeled as N and S. The source S is marginally resolved, showing a slight elongation to the south. Crosses mark the positions of the two components S1 and S2 detected at 7 mm. At 7 mm, the source N is detected as a weak source, and the 1.3 cm source S, is resolved in two components, S1 and S2. }}

\label{FigXZ}  
\end{center} 
\end{figure*}

\begin{deluxetable*}{ccccccccc}
\tabletypesize{\scriptsize}
%\rotate
\tablewidth{0pt}
%\tablenum{}
\tablecaption{Components of the XZ~Tau triple system\label{TabXZ}}
%\tablehead{}
%\tablecolumns{}
\startdata
\hline \hline 
 Radio    & Optical	   	      & \multicolumn{3}{c}{Position (J2000)\tablenotemark{a}}      & \multicolumn{2}{c}{Flux Density (mJy)\tablenotemark{a}} &  Spectral       &  Emission	\\ \cline{3-5} \cline{6-7}
Component & Component      	      &    RA     &	 DEC   & Error &	 1.3 cm         &       7 mm	     &  Index	       &  Mechanism	\\ \hline
%----------------------------------------------------------------------------------------------------------------------------------------------------------------------------------
 N        & XZ~Tau~B	   	      &  04 31 40.0751  &  18 13 57.079     &  0$\farcs$006   &  0.14 $\pm$ 0.03  &  0.3 $\pm$ 0.2 &  1.1 $\pm$ 1.0  &  Free-free        \\
 S1       & XZ~Tau~A	   	      &  04 31 40.0891  &  18 13 56.853     &  0$\farcs$005   &  0.33 $\pm$ 0.03\tablenotemark{b}  &  1.2 $\pm$ 0.2 &  1.9 $\pm$ 0.3  &  Free-Free + Dust        \\
 S2       & XZ~Tau~C\tablenotemark{c} &  04 31 40.0899  &  18 13 56.762     &  0$\farcs$008   &  0.08 $\pm$ 0.03\tablenotemark{b} &  0.8 $\pm$ 0.2 &  3.4 $\pm$ 0.7  &	 Dust  \\ \hline
%----------------------------------------------------------------------------------------------------------------------------------------------------------------------------------
\enddata 

\tablenotetext{}{\textsc{Notes.-} Units of right ascension are hours, minutes, and seconds, and units of declination are degrees, arcminutes, and arcseconds.\\}
\tablenotetext{a}{Derived from Gaussian fits to the 7 mm natural weighting map (Epoch 2004.8). Errors in positions correspond to the relative positions. The error in the absolute position is estimated to be $\sim$0$\farcs$01.}
\tablenotetext{b}{Flux density derived by fitting the marginally resolved 1.3 cm source S to two unresolved Gaussian sources.}
\tablenotetext{c}{This component is probably a deeply embedded object undetectable at optical/IR wavelengths.}

\end{deluxetable*}

\section{Results and discussion}

\subsection{HL~Tau}

 In Figure \ref{FigHL}a we show a superposition of the 7 mm emission of HL~Tau (contours) over the 1.3 cm map (colors). At 1.3 cm, HL~Tau shows a quadrupolar morphology consisting of two nearly perpendicular structures at position angles (P.A.s) of $\sim$50$^\circ$ and $\sim$140$^\circ$. As previously noted by Greaves et al. (2008), the alignment of the 1.3 cm emission at P.A.$\simeq$50$^\circ$ with the optical jet of HL~Tau (e.g. Anglada et al. 2007), suggests that the 1.3 cm emission at this P.A. is tracing free-free emission from the shock-ionized gas at the base of the flow. In contrast, the 1.3 cm emission at P.A.$\simeq$140$^\circ$ is most probably tracing thermal dust emission from a circumstellar disk perpendicular to the outflow. This is more evident in our 7 mm map (see Fig. \ref{FigHL}a), where the emission shows an elongated morphology with a P.A. of $\sim$140$^\circ$ (perpendicular to the jet) and a size of $\sim$50 AU. The positions of the 1.3 cm and 7 mm maxima are nearly coincident and define the center of the radio jet. 
 
 We also detected two weak compact 1.3 cm features, one $\sim$0$\farcs$4 to the NW (at a level of 3-$\sigma$) and the other $\sim$0$\farcs$4 to the S (at a level of 4-$\sigma$) of the 1.3 cm emission peak (see Fig. \ref{FigHL}a). Greaves et al. (2008) identified the 1.3 cm NW feature with a 1.4 mm source detected by Welch et al. (2004), obtaining a spectral index of 2.5 in this wavelength range. They proposed that the 1.3 cm NW feature is dust emission arising from a proto-planet with a mass of 14 M$_J$ and orbiting around HL~Tau at a radius of $\sim$65 AU. The other 1.3 cm feature to the S, although somewhat stronger than the NW feature, was not discussed by these authors. As can be seen in Figure \ref{FigHL}a, we have not detected 7 mm emission at the position of the proto-planet candidate, with a 3-$\sigma$ upper limit level of $\sim$0.15 mJy. This upper limit and the 1.3 cm flux density of Greaves et al. (2008) implies a spectral index $<$1, suggesting free-free rather than dust emission. On the other hand, the spectral index of 2.5 between 1.3 cm and 1.4 mm obtained by Greaves et al. (2008) would imply a 7 mm flux density of 0.5 mJy, well above our 3-$\sigma$ upper limit. Additional, high angular resolution observations in the mm range are necessary to clarify the true nature of this source. 
  
 In Figure \ref{FigHL}b we show an intensity profile of the 7 mm emission along the major axis of the disk. As can be seen in Figures \ref{FigHL}a and \ref{FigHL}b, the 7 mm emission is knotty and asymmetric. The proposed disk extends $\sim$25 AU to the SE, and shows a gap at a radius of $\sim$10 AU and the emission to the NW appears truncated at a radius of $\sim$12-15 AU. The gap and the truncation are probably related to a decrease of the density in the disk. Thus, since the disk is observed nearly edge-on (e.g. Cohen 1983), this gap could be associated with an annulus of low density in the disk with a radius of $\sim$10-15 AU, similar to the radius of the orbit of Saturn, or twice the orbit of Jupiter. Such structures are expected to form as a consequence of the accretion of a planetary object forming in the disk (see, e.g. Bryden et al. 1999). Therefore, the gap and the truncation of the disk detected in our 7 mm map could be related to the formation of a proto-planet. Alternatively, this gap could be due to tidal truncation in a binary system (e.g., Pichardo et al. 2005) as in the case of HH30 (Anglada et al. 2007, Guilloteau et al. 2008). However, with the present data, these suggestions are very speculative, and should be confirmed by subsequent observations and/or modelling of the disk of HL~Tau.

\subsection{XZ~Tau}
\begin{figure*}  
\begin{center}

\includegraphics[width=\textwidth]{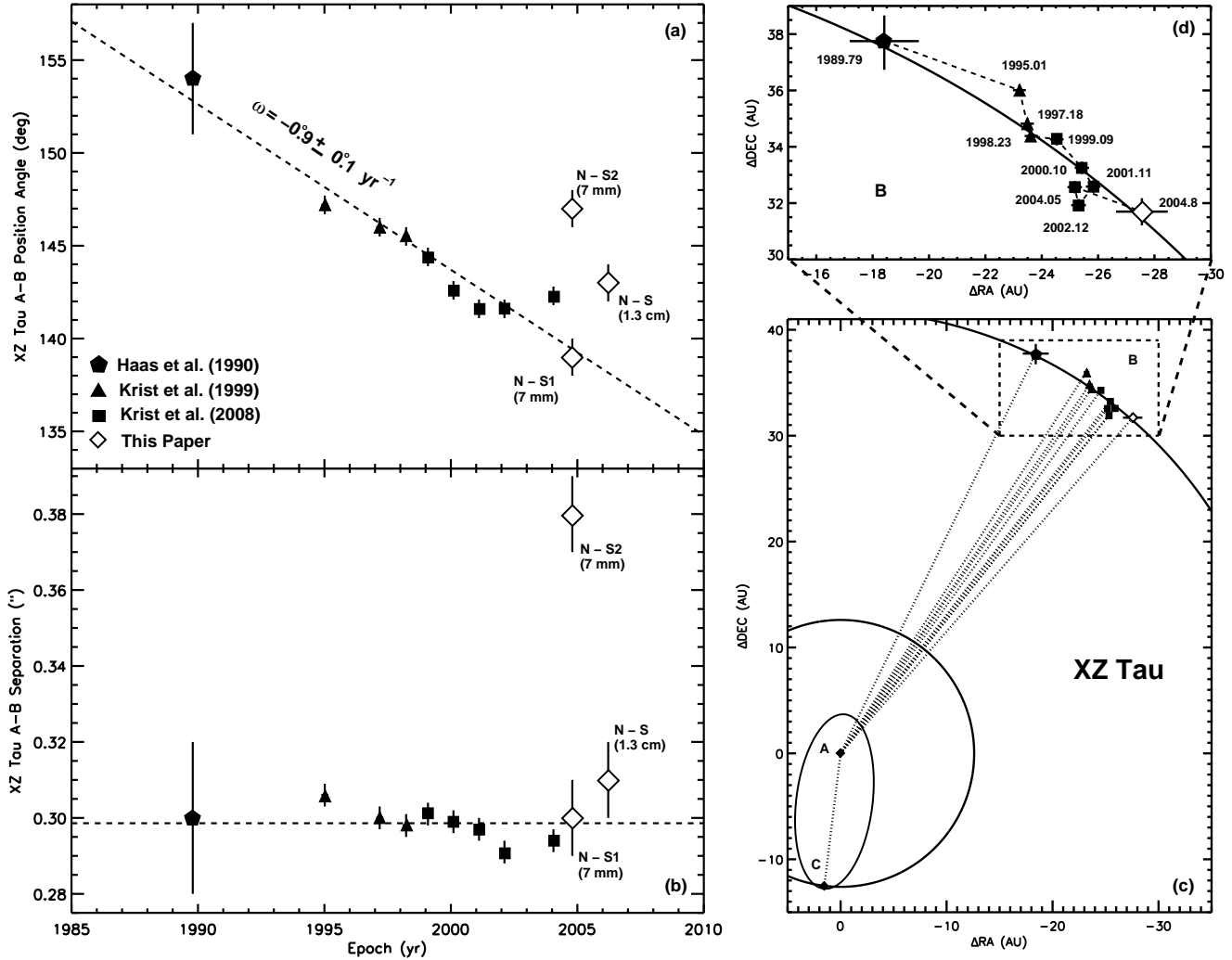}

\caption{\footnotesize{\textbf{(a)} Plot of the P.A. between the optical/IR components XZ~Tau~A and XZ~Tau~B vs. time (Haas et al. 1990; Krist et al 1999, 2008). The dashed line is a least squares fit to these data, and corresponds to an orbital angular velocity of $-$0.9 $\pm$ 0.1 degrees yr$^{-1}$. Three different P.A.s measured from our 1.3 cm and 7 mm maps are marked with diamonds. Only the P.A. between the 7 mm sources N and S1 seems to be compatible with the angular orbital velocity derived from the optical/IR data. This suggests that the source S1 is the radio counterpart of the optical component XZ~Tau~A. \textbf{(b)} Plot of the separation between the optical/IR components XZ~Tau~A and XZ~Tau~B vs time. The measurements are compatible with a constant separation of 0$\farcs$30$\pm$0$\farcs$01. \textbf{(c)} and \textbf{(d)} A schematic representation of the XZ~Tau triple system. The positions of XZ~Tau~B relative to XZ~Tau~A, derived from the P.A. and separation measurements of  Haas et al. (1990), Krist et al (1999, 2008) and our data, are shown. The current data are compatible with a circular face-on orbit of XZ~Tau~B relative to XZ~Tau~A. In contrast, with only a single detection of the third component, XZ~Tau~C, we have no information about its orbital motion. Two hypothetical orbits (circular and elliptical) for XZ~Tau~C are shown.}}

\label{FigOrb}  
\end{center}
\end{figure*}

 In Figure \ref{FigXZ}a we show the 3.6 cm continuum map (angular resolution $\simeq$0$\farcs$3) of XZ~Tau, obtained by Rodr\'{i}guez et al. (1994), where a single radio continuum source was detected. Figures \ref{FigXZ}b and \ref{FigXZ}c show our 1.3 cm and 7 mm maps, respectively. At 1.3 cm (angular resolution $\simeq$0$\farcs$1), XZ~Tau is resolved in two radio continuum sources (labeled N and S) separated by $\sim$0$\farcs$3 with a P.A. of $\sim$143$^\circ$. The 7 mm map (angular resolution $\simeq$0$\farcs$05), reveals that the 1.3 cm component S is actually a double radio source (that we label as components S1 and S2) with a separation of 0$\farcs$09 and a P.A. of $-$6$^\circ$. 

 The size of the synthesized beam of the 1.3 cm map ($\sim$0$\farcs$1), slightly larger than the separation between the S1 and S2 components at 7 mm ($\sim$0$\farcs$09), does not allow to completely separate the emission of these components at 1.3 cm. However, the 1.3 cm source S is marginally resolved in our map, showing a slight elongation to the south. In Figure \ref{FigXZ}b we show the positions of the components S1 and S2 derived from the 7 mm map. As can be seen, most of the 1.3 cm emission seems to arise from component S1, since this is the source nearest to the peak of the 1.3 cm source S. The 1.3 cm source S can be fitted by two unresolved Gaussian sources separated by $\sim$0$\farcs$08, very similar to the separation between sources S1 and S2 in the 7 mm map. 

 In Table \ref{TabXZ} we give the position (derived from the 7 mm map), flux density at 1.3 cm and 7 mm, and the spectral index for each component. For source N, we obtained a spectral index of 1.1 $\pm$ 1.0 consistent with free-free emission from this source. For source S as a whole, we obtained a ``combined'' spectral index of 2.5 $\pm$ 0.3 (obtained from the total flux density of the 1.3 cm source S, and the sum of the 7 mm flux densities of S1 and S2). This steep spectral index suggests a combination of free-free plus thermal dust emission. If we use the flux densities obtained from a double Gaussian fit to the 1.3 cm source S, for source S1 we obtain a spectral index of 1.9 $\pm$ 0.3 which suggests a combination of free-free plus thermal dust emission. In contrast, for S2 we obtain a spectral index of 3.4 $\pm$ 0.7 which suggests that the emission of this source is dominated by thermal dust emission. 

 As mentioned above, XZ~Tau is known to be an optical/IR binary system with a separation of $\sim$0$\farcs$3. Following the notation of Krist et al. (2008), we will refer to the optical components of this system as XZ~Tau~A (the southern optical component) and XZ~Tau~B (the northern optical component). Krist et al. (1999, 2008) observed XZ~Tau with the HST in several epochs, and detected a change in the P.A. of the binary due to orbital motions. According to this motion ($\sim -$0.9 deg yr$^{-1}$) at the epoch of our radio observations, component XZ~Tau~B must be still located north of component XZ~Tau~A. Therefore, we identify the radio source N with the optical component XZ~Tau~B. 

 The identification of the two radio sources at the south (S1 and S2) with the other optical component (XZ~Tau~A) is not straightforward. One possibility is that both radio sources have optical emission that could not be resolved by the HST observations of Krist et al. (2008). However, another possibility is that only one of the radio sources is the counterpart of the optical component. In order to investigate these possibilities, we have plotted in Figures \ref{FigOrb}a and \ref{FigOrb}b the P.A. and separation of the optical binary measured in each epoch together with the P.A.s and separations between the different radio components measured from our 1.3 cm and 7 mm maps. As can be seen, the P.A. and separation between the radio sources N and S1 seems to be the only one compatible with the previous optical/IR data. This suggests radio source S1 as the sole counterpart to the optical component XZ~Tau~A; S2 is optically undetected. This is similar to the case of the close binary SVS~13, where both components are detected at cm wavelengths, but only one object is detected in the optical regime (Anglada et al. 2004). As we have discussed above, the radio emission of component S2 seems to be dominated by thermal dust emission. We believe that this source traces a star since if it not were gravitationally bound, it would quickly dissipate. The size of Sb is smaller than 7 AU, that expanding at only 1~km~s$^{-1}$, will dissipate in only 30 years. Then, we conclude that this dust must be gravitationally bound to a star and it is most possibly an envelope or a disk. This suggests that source S2 is a more deeply embedded object so it is not surprising that it is optically very faint. Therefore, we propose that the radio source S2 is an optically invisible third component of the XZ~Tau system. For consistency with the previous nomenclature, we call this component XZ~Tau~C. 
  
 In summary, the results derived from our 1.3 cm and 7 mm maps suggest that XZ~Tau, previously known as a binary system, is actually a triple star-system (see Figs. \ref{FigOrb}c and \ref{FigOrb}d). The previously known optical/IR components, XZ~Tau~A and XZ~Tau~B, have 1.3 cm and 7 mm counterparts and are separated by 42 AU. A third component, XZ~Tau~C has been detected for the first time in our 1.3 cm and 7 mm maps, and is found to be a more deeply embedded object, probably undetectable at optical/IR wavelengths, that forms a close binary system with XZ~Tau~A with a separation of only 13 AU. 

 Each optical component of the XZ~Tau system (A and B) is known to drive a collimated jet (Krist et al 2008). It has been proposed that the jet of XZ~Tau~A underwent a large velocity pulse approximately in 1980 (Krist et al. 2008) which, after shocking and heating the gas near the star, would result in the 70 km s$^{-1}$ expanding bubble detected in the series of HST images of Krist et al. (1999, 2008). These authors explored the possibility that this ejection was triggered by the passage of component XZ~Tau~B very near component XZ~Tau~A. However, even assuming a very eccentric orbit (e$\simeq$0.9) they derived that the last approach of these components would have taken place in 1955, and thus, they discarded this possibility.

 The discovery of a new companion, closer to XZ~Tau~A, suggests that it is once more feasible to consider the passage of this companion as the ejection trigger. If we assume a face-on, nearly circular (e$\simeq$0.1) orbit and a total mass of 1 M$_\sun$ (as suggested by the results of Krist et al. 1999 and White \& Ghez 2001), we derive an orbital period of $\sim$40 yr. If at the time of our 7 mm observations (2004.8) the XZ~Tau~A-C system wass near apoastron, then the last periastron passage must have taken place one half-period before, i.e. nearly 1980, the date estimated for the ejection. Further observations are needed to produce a more accurate ephemeris for the XZ Tau system. The present results predict a new periastron passage around 2020, and then a new major ejection from XZ~Tau~A at this epoch.

\emph{Acknowledgements.} We thank A.M.S. Richards for valuable comments that improved the paper. C.C.-G. acknowledges support from MEC (Spain) FPU fellowship. G.A. and C.C.-G. acknowledge support from MEC (Spain) AYA2005-08523-C03-03 and MICINN (Spain) AYA2008-06189-C03-01 grants (co-funded with FEDER funds), and from Junta de Andaluc\'{\i}a (Spain). L.F.R. and S.C. acknowledge the support of DGAPA, UNAM, and CONACyT (M\'exico).

\newpage

%TABLES

%FIGURES

\end{document}